# Using qqZ Events
# to "Calibrate" Vector Boson Fusion at the LHC

Dan Green
Fermilab
February, 2005

## Abstract

Vector Boson Fusion (VBF) is a promising discovery level process in the Higgs search at the LHC [1]. Nevertheless, that search depends on understanding the detector response to a good level of accuracy. Therefore, it is useful to have a known process by which to validate the search methodology. The VBF production of a Z with subsequent leptonic decay appears to be ideal in that the Feynman diagrams are the same for Z and Higgs VBF production and there is a clean resonance in the dilepton spectrum. In addition, the cross section for Z is larger than that for light Higgs production so that the calibration process can be studied first. Because the masses of the Z and a light Higgs are similar, the "tag jet" kinematics is also very similar in the two cases. Because the VBF process has not yet been observed and isolated at the Tevatron background processes must be well evaluated and then removed or reduced.

## ggZ Background

The inclusive production of Z bosons is well measured at the Fermilab Tevatron and the cross section can be reliably extrapolated to the LHC [2]. Estimates for the production of Z plus one jet and Z plus two jets have also been made in Ref.2 and elsewhere.

As a rough check the COMPHEP program [3] was used to calculate the cross sections for Z, Z+g and Z+g+g events. The results were found to be near those quoted in Ref.2, which validates the event generation at the tree level. In fact, of the many Feynman diagrams, the relevant ones producing the largest cross section in the Z+g+g process, called (ggZ) in what follows, are shown below in Fig.1. Basically, the Drell-Yan production of a single Z boson is augmented by initial state radiation (ISR) of two gluons or by one gluon with subsequent "splitting" of that gluon into a gluon pair.

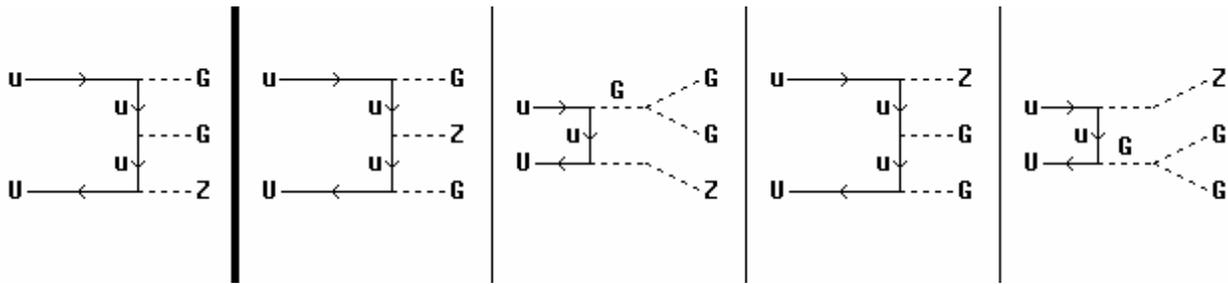

Figure 1: Feynman diagrams for Drell-Yan production of a Z boson and two radiated gluons in quark- antiquark annihilations, called (ggZ).

## qqZ – VBF and QCD, gg Backgrounds

There are additional processes leading to a Z boson plus two jets in the final state. The VBF process is shown in Fig.2 below. The VBF process is the radiation of a virtual gauge boson (only W here as opposed to H production because the ZZZ vertex is absent in the Standard Model, while the ZZH vertex is allowed) by a valence quark from both incoming protons after which the gauge bosons fuse to create a boson, either the scalar Higgs or the vector Z. Note that the (ggZ) process has a diagram with similar topology to the VBF process, called (qqZ)$_{VBF}$ in what follows.

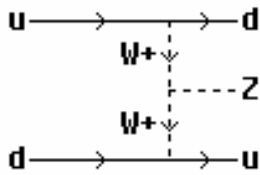

Figure 2: Feynman diagram for the VBF production of a Z boson by valence quarks in the incoming protons.

However, in addition to the (qqZ)$_{VBF}$ process there are also QCD processes involving valence quarks in the initial state, which make a strong scattering by way of gluon exchange . The Z is produced by ISR off the incoming quarks or final state radiation (FSR) off the outgoing quarks. Typically, there are 17 Feynman diagrams describing this process. However, the ISR and FSR diagrams are the only major contributors to the cross section and they are shown in Fig.3. In what follows only these four diagrams are used in order to speed up the calculation. This process is called (qqZ)$_{QCD}$ in what follows.

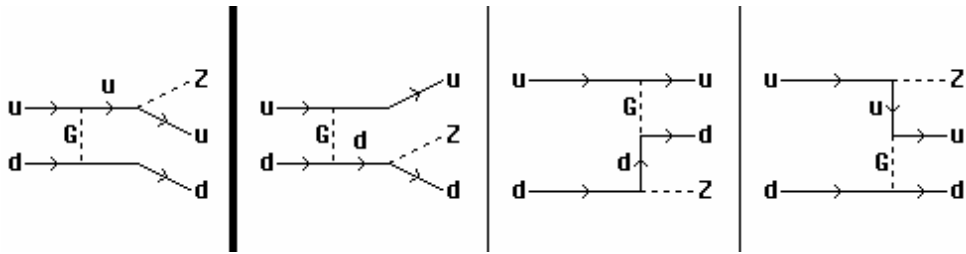

Figure 3: Feynman diagrams for the strong scattering of valence quarks accompanied by the ISR or FSR of a Z boson by the incoming or outgoing quarks.

Finally, there are processes where a gluon pair fuses to emit a Z and a quark pair, called (qqZ)$_{gg}$ in what follows. The relevant diagrams with an up quark pair in the final state are shown in Fig.4.

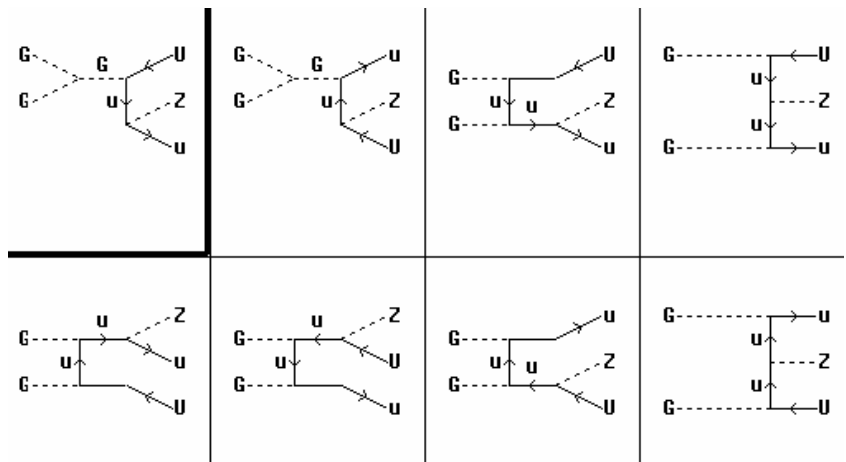

Figure 4: Feynman diagrams for the gluon-gluon production of a Z boson plus a quark pair.

## gqZ Background

In addition to the Z plus gluons and Z plus quarks final states there is a background with a Z plus one quark and one gluon. The eight Feynman diagrams for the case of an up quark in the initial state are shown in Fig.5. Note that there is a diagram with the VBF topology. This process is called (gqZ) in what follows.

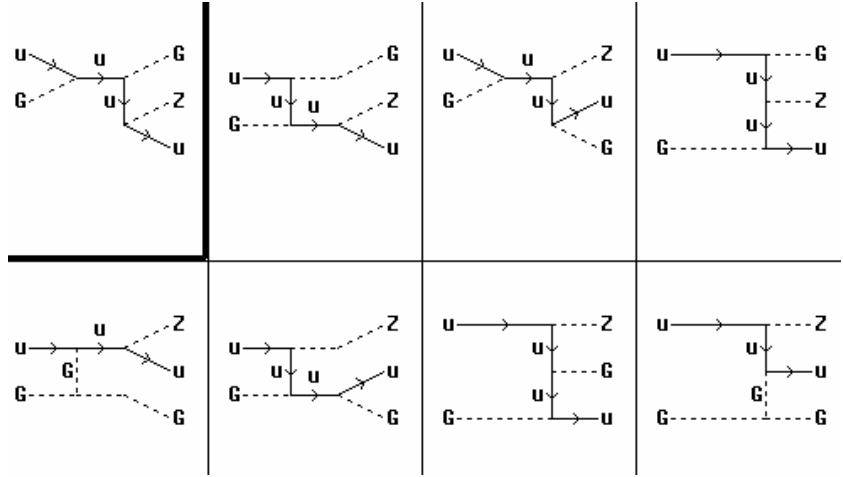

Figure 5: Feynman diagrams for the production of a Z plus quark plus gluon in the final state.

## COMPHEP Generation and Cut Studies

The four classes of processes shown above are the major backgrounds to the measurement of the VBF production of a Z boson at the LHC. Other processes which were examined either have smaller cross sections or are more easily removed by the cuts applied as described below. Events were typically generated without kinematic cuts. However, in some cases a transverse momentum cut of 5 GeV was applied to an outgoing particle if there was a radiative pole present in the process. The generated cross section is shown in Table 1.

The process with the largest uncut cross section is the (gqZ) process. However, the background most resistant to removal will turn out to be the strong scattering of valence quarks accompanied by radiation of a Z, the $(qqZ)_{QCD}$ process. Ultimately, that difficulty arises from the fact that the strong scattering of quarks causes forward backward emission of outgoing quarks and that mimics the "tag jets" shown in Fig.2.

The COMPHEP program was used to make an initial study at the simple tree level and generator level to see if the VBF process can be extracted from the backgrounds. All processes were generated in COMPHEP with the requirement that the transverse momentum of all final state particles was greater than 5 GeV when needed. The cross sections for a set of imposed cuts is shown in Table 1 where the cuts are imposed sequentially. A generic parton is assumed to create a final state jet, or (jjZ) final state.

**Table 1**
**Cross Sections for VBF and**
**Background Processes in jjZ Production**

| Cut | ggZ (pb) | (gqZ)$_{gq}$ | (qqZ)$_{QCD}$ | (qqZ)$_{gg}$ | (qqZ)$_{VBF}$ |
|---|---|---|---|---|---|
| none | 3000 | 20,000 | 340 | 1700 | 13.4 |
| y$_1$=(-5,-1) & y$_2$=(1,5) | 101 | 860 | 150 | 190 | 10.2 |
| M$_{12}$ > 750 GeV | 2.3 | 4.3 | 27.0 | 0.88 | 5.8 |
| y$_1$-y$_Z$>1.5 & y$_Z$-y$_2$>1.5 | 0.45 | 2.8 | 1.85 | 0.41 | 2.9 |
| P$_{TZ}$ > 80 | 0.15 | 1.3 | 0.91 | 0.20 | 2.4 |
| Other cuts? M$_{1Z}$,M$_{2Z}$ Harder on angular ordering | | 606, 403 | 612, 544 GeV | | 668, 643 GeV |

Initially, the VBF process is buried by a factor of about 2000. The first cut simply requires two "tag jets', one in the forward hemisphere and one in the backward. In making rapidity cuts what was done was to assume massless partons and to cut on the pseudorapidity. Clearly, this has a small effect of the VBF process, where the mean rapidity of the tag jet is ~ 1.9. However, this cut improves the signal to background ratio by about a factor of ten. The scatter plots of the rapidity of one tag jet, y$_1$, with respect to the other, y$_2$ for the five processes is shown in Fig.6. Note the VBF topology for the qqZ events, with y$_1$ ~ - y$_2$.

For the backgrounds, typically the two jets in the final state have similar rapidities, y$_1$ ~ y$_2$ while the VBF signal has opposite rapidities. The exception is that for (qqZ)$_{QCD}$ events which are not strongly removed by the tag jet cut. Note also the strong correlations, y$_1$ ~ y$_2$ in the case where a gluon splitting (see Fig.1) occurs. The azimuthal correlation of the two jets was also studied. Although there was an initial difference between the signal and background events, it largely disappeared after the cuts were imposed. Therefore, this cut was not imposed.

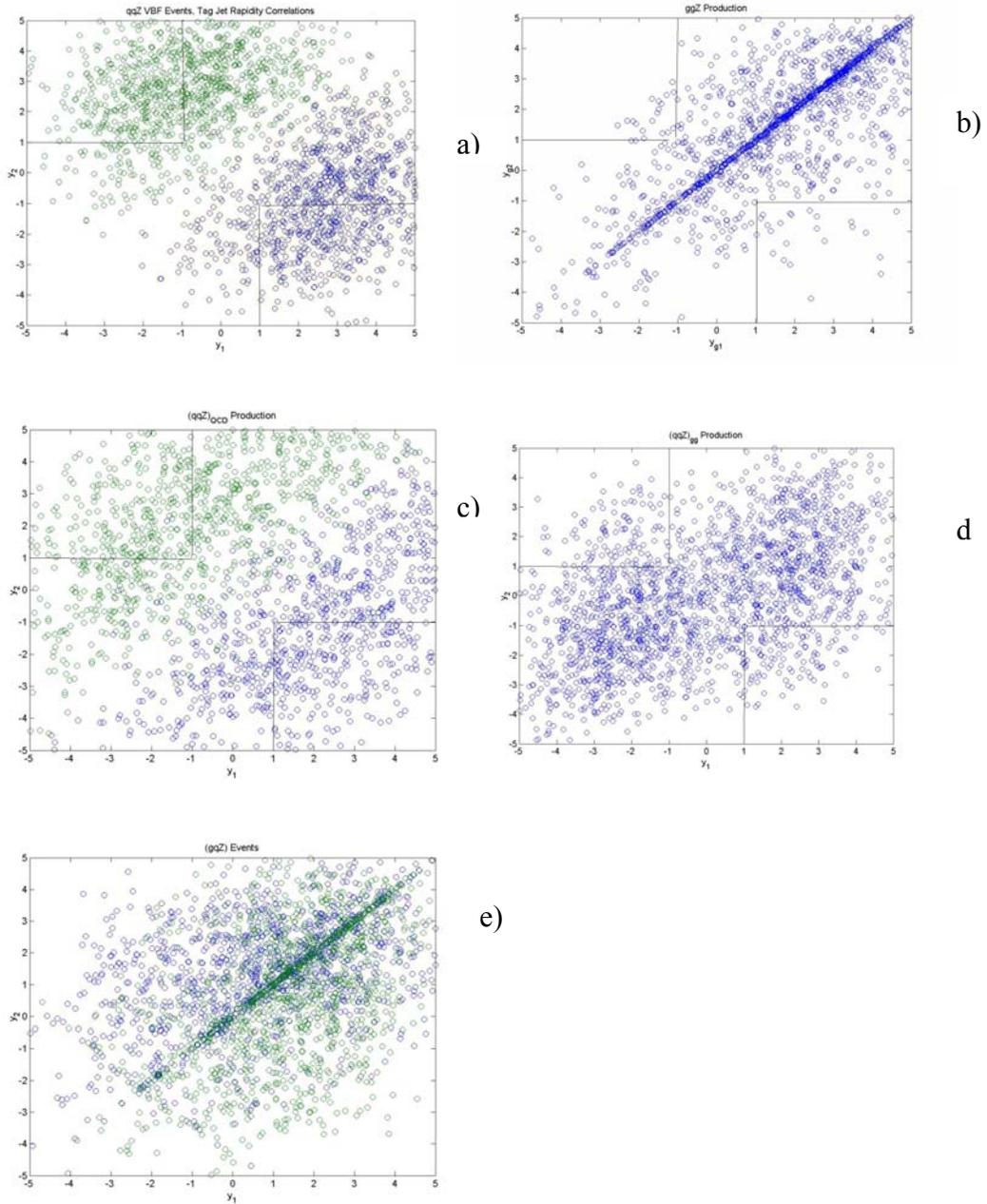

Figure 6: Correlation of the two "tag jets" in the jjZ final state for the five processes considered in this note: a)- $(qqZ)_{VBF}$, b)-(ggZ), c)- $(qqZ)_{QCD}$, d)- $(qqZ)_{gg}$, e) -(gqZ). The lines indicate the regions of the "tag jet" cuts.

The second cut indicated in Table 1 is on the mass of the two tag jets. The tag jets, because the radiation of the virtual W's is "soft", tend to follow the kinematics of the initial state partons [4]. Therefore, a cut on the minimum value of the mass of the two tag jets is a very useful requirement. Note that subsequent distributions are shown after all prior cuts have been imposed. In this case the cut is on the mass defined to be, $M_{12}^2 = (p_1 + p_2)^2 - (\vec{p}_1 + \vec{p}_2)_{\parallel}^2$, indicating that only the longitudinal components of the momentum are subtracted from the energy. The distributions for the five processes are shown in Fig.7. The mean value of the tag jet mass was found to be for the (qqZ)$_{VBF}$ process, 1013 GeV, for the (ggZ) process, 162 GeV, for (gqZ) events 277 GeV, for (qqZ)$_{gg}$ events 177 GeV, while for the (qqZ)gg process the mean mass is 462 GeV.

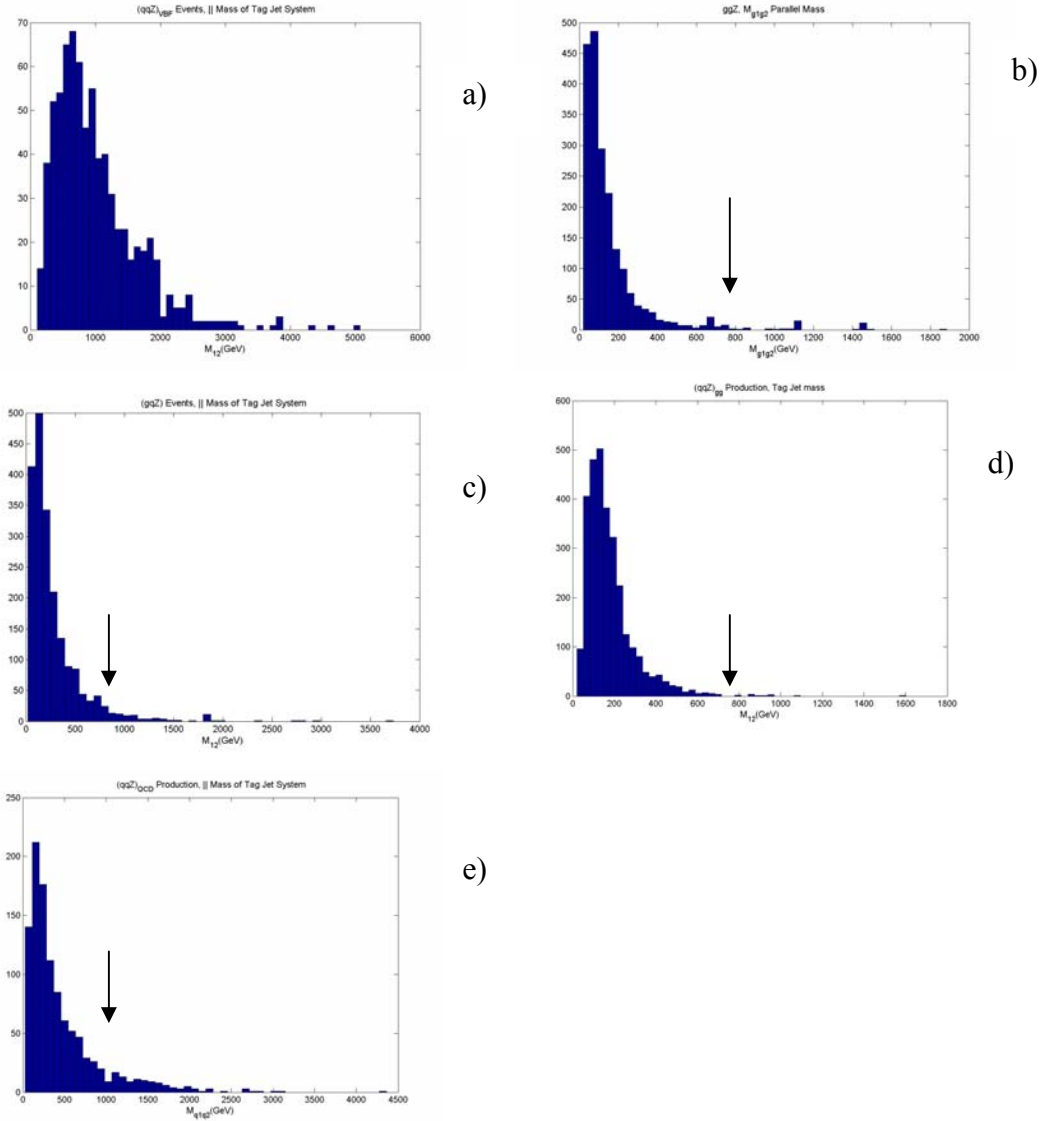

Figure 7: Mass of the tag jet pairs for the five processes considered in this paper, a)-(qqZ)$_{VBF}$, b)-(ggZ), c)-(gqZ), d)-(qqZ)$_{gg}$, e)-(qqZ)$_{QCD}$. The mass is computed subtracting only the longitudinal momentum from the squared energy. The arrows indicate the cut value used in this study.

The mass cut on the tag jets improves the signal to noise ratio by about another factor of one hundred except in the case of the $(qqZ)_{QCD}$ events. After making this cut the signal to background ratio is about one and the major background is now the $(qqZ)_{QCD}$ process. In turn, this background can be reduced by requiring an "angular ordering". In the VBF process the produced Z is in the middle of phase space, well separated from the tag jets, see Fig.2. In the other processes, this angular ordering is not obviously respected. Therefore, the cut that the Z leads the negative rapidity tag jet by 1.5 in rapidity and follows the positive rapidity tag by 1.5 was made. The distributions are shown in Fig.8, where the lines indicate the cuts imposed.

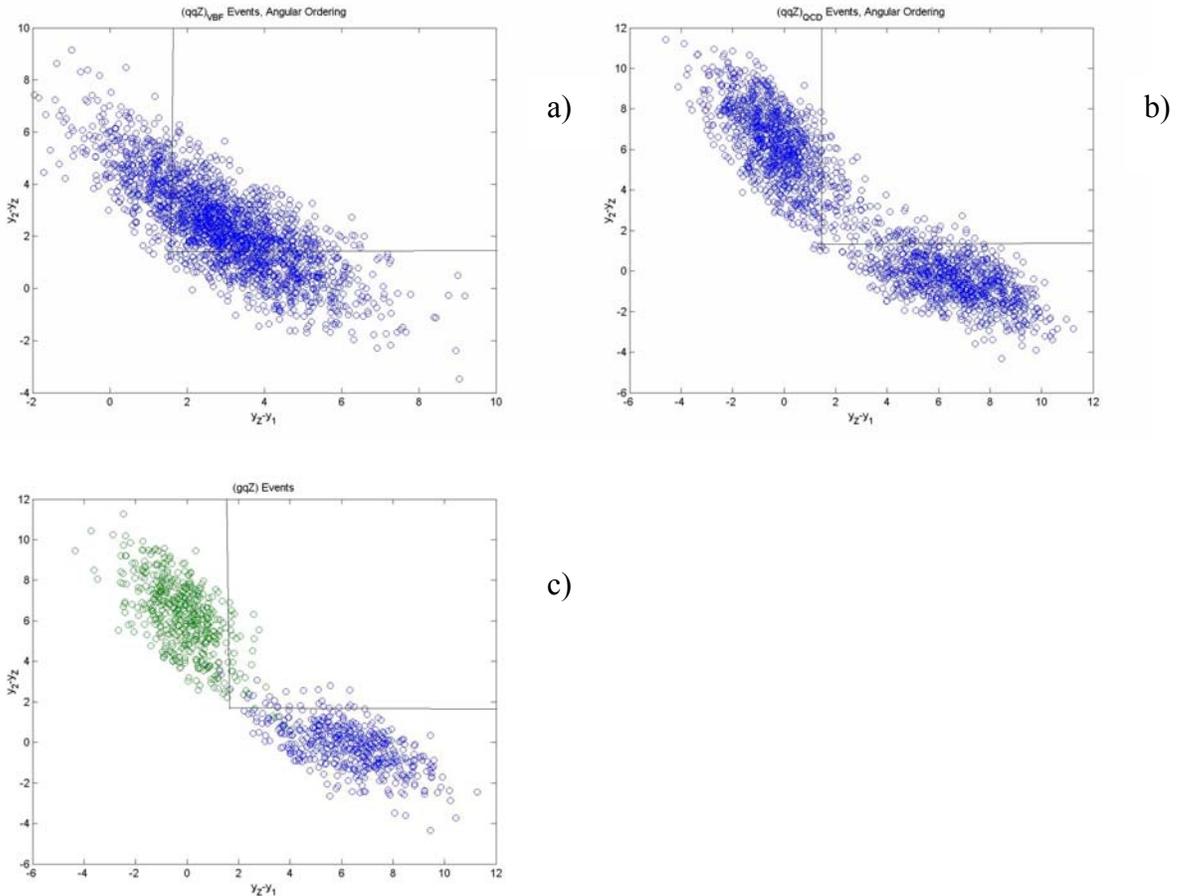

Figure 8: Angular ordering in rapidity between the Z and the two tag jets for: a) - $(qqZ)_{VBF}$ events, b) - $(qqZ)_{QCD}$ events and (gqZ) events. The cuts are indicated by the lines shown on the scatter plots.

Clearly this cut is quite effective in removing the remaining large backgrounds and it could be made more stringent. However, the signal events are being substantially removed and the signal phase space is being distorted. After making this cut the signal VBF events exceed those of any of the four backgrounds considered here. In order to clean up the signal more we use the fact that the Z from the backgrounds is radiated by low mass partons, while the VBF produced Z boson are given a substantial transverse momentum in the production process. The transverse momentum distribution of the Z bosons are shown in Fig. 9. In the case of the VBF signal, the mean is 168 GeV. For the $(qqZ)_{QCD}$ events, the mean is lower, 101 GeV, as expected. For the (gqZ) events, the mean is also lower, 96 GeV in this case.

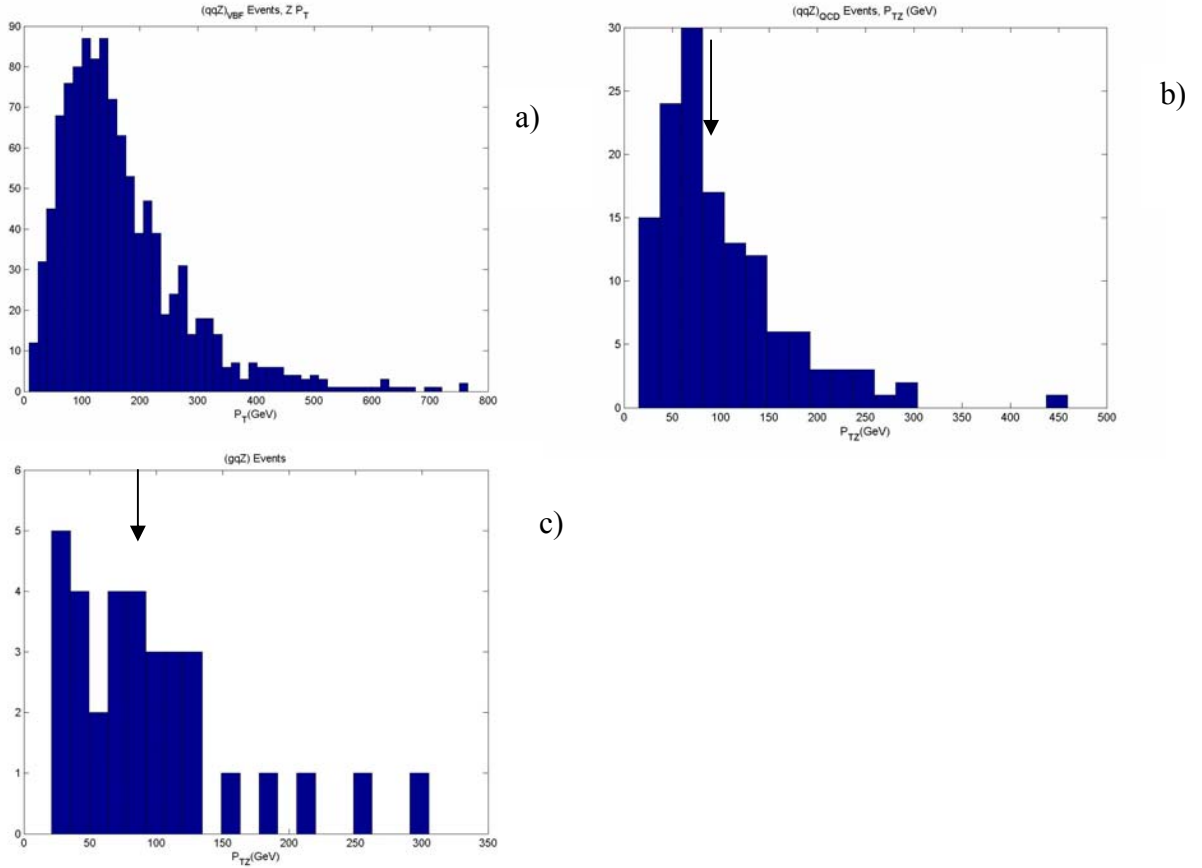

Figure 9: Transverse momentum of the Z bosons produced in; a) - (qqZ) $_{VBF}$, b) - $(qqZ)_{QCD}$ and c) – (gqZ) processes after all prior cuts have been imposed. The cut at 80 GeV is indicated by the arrows.

## Conclusions

After making this fourth cut the signal is roughly equal to the sum of the backgrounds. Clearly, more stringent angular ordering cuts can be made to improve the signal to background. Cuts are also possible on the mass of the Z and tag jets, as indicated in Table 1 in the last row. However, the large backgrounds have forced us to make a series of cuts of some severity. After making sufficient cuts so that the signal to background ratio is of order one, the signal itself has been reduced by a factor of about six. These cuts have also made the signal and background kinematically rather similar.

Therefore, although we have managed to make to VBF process manifest, it is now observed as a 'counting experiment". That fact implies that a full confidence that one has observed the VBF process at the LHC requires that the efficiency of the detectors in triggering and reconstruction be rather well understood.